\documentclass[rnote,traditabstract]{aa}
\usepackage{graphicx}
\usepackage{txfonts}
\usepackage{rotating}
\usepackage{lscape}
\usepackage{longtable}
\usepackage{natbib}
\begin{document}

\title{Flat-spectrum radio sources as likely counterparts of unidentified INTEGRAL sources}
\titlerunning{Flat spectrum radio sources as counterparts of INTEGRAL sources}
\authorrunning{M.~Molina}
\author{M. Molina\inst{1} \and R. Landi\inst{1} \and L. Bassani\inst{1} \and A. Malizia\inst{1} 
\and J.~B Stephen\inst{1} \and  A. Bazzano\inst{2} \and A.~J. Bird\inst{3} \and N. Gehrels\inst{4}
 }
 
\offprints{molina@iasfbo.inaf.it}
\institute{IASF/INAF, via Gobetti 101, I-40129 Bologna, Italy \and
IAPS/INAF, Via del Fosso del Cavaliere 100, 00133 Rome, Italy \and
School of Physics and Astronomy, University of Southampton,
        SO17 1BJ, Southampton, U.K. \and
NASA Goddard Space Flight Center, Greenbelt, MD 20771        
        }

\date{Received  / accepted}

      
\abstract
{Many sources in the 4$^{\rm th}$ INTEGRAL/IBIS catalogue are still 
unidentified since they lack an optical counterpart.
An important tool that can help in identifying and classifying 
these sources is the cross-correlation with radio catalogues, 
which are very sensitive and positionally accurate. Moreover, the radio properties of a source, 
such as the spectrum or morphology, could provide further 
insight into its nature. In particular, flat-spectrum radio sources at high Galactic 
latitudes are likely to be AGN, possibly associated to a blazar 
or to the compact core of a radio galaxy. Here we present a small sample 
of 6 sources extracted from the 4$^{\rm th}$ INTEGRAL/IBIS
catalogue that are still unidentified and/or unclassified,
but which are very likely associated with a bright, flat-spectrum radio object.
To confirm the association and to study the source
X-ray spectral parameters, we performed X-ray follow-up observations with
Swift/XRT of all objects.
We report in this note the overall results obtained from this
search and discuss the nature of each individual INTEGRAL source.
We find that 5 of the 6 radio associations are also detected in X-rays; 
furthermore, in 3 cases they are the only counterpart found. 
More specifically, IGR J06073--0024 is a flat-spectrum radio quasar at z=1.08, 
IGR J14488--4008 is a newly discovered radio galaxy, while IGR J18129--0649
is an AGN of a still unknown type.
The nature of two sources (IGR J07225--3810 and IGR J19386--4653) 
is less well defined, since in both cases we find another X-ray source in the INTEGRAL error circle;
nevertheless, the flat-spectrum radio source, likely to be a radio loud AGN,  
remains a viable and, in fact,a more  convincing association in both cases. 
Only for the last object (IGR J11544--7618) could we not find any convincing counterpart  
since the radio association is not an X-ray emitter, while the only
X-ray source seen in the field is a G star and therefore unlikely to produce the 
persistent emission seen  by INTEGRAL.

}

\maketitle

\section{Introduction}

One of the key objectives of the INTEGRAL mission \citep{Winkler:2003} 
is to survey the sky at high energies ($>$20\,keV), where non-thermal processes 
take place, the effects of absorption are drastically reduced, and  
extreme astrophysical phenomena are observed. This survey is performed by taking advantage of the
unique imaging capabilities of the IBIS instrument \citep{Ubertini:2003}, which allows
the detection of sources at the mCrab flux level, with an angular resolution of 12\arcmin and a
point source location accuracy of typically 1--3\arcmin within a large (29$\times$29 degrees)
field of view.

Up to now, several surveys have been compiled from data collected by IBIS, the largest 
one being the fourth catalogue by \citet{Bird:2010}, which
lists 723 hard X-ray sources of both Galactic and extragalactic nature. 
However, many objects in this survey
are still unidentified, i.e. they have no obvious counterpart in other wavebands 
and so cannot be properly
classified. Their classification remains one of the main objectives
of the survey work, but it is made difficult by the large IBIS positional uncertainty.
For this reason, accurate arcsecond localisation is pivotal to
pinpoint their likely optical counterpart and assess their nature
by means of spectroscopic follow-up observations (e.g. \citealt{masetti12}
and references therein). Furthermore, many of these unidentified INTEGRAL 
sources are newly discovered and therefore lack multiwaveband data, 
which could provide important clues to their overall properties, hence nature.

An important tool that can help in identifying these sources is the cross-correlation
with catalogues in other wavebands, for example in the radio. There are many valid 
reasons to use radio catalogues for counterpart searches:   
some high-energy emitting objects are also radio sources, such as active galactic nuclei (AGN),
pulsars/pulsar wind nebulae, and X-ray binaries; radio surveys are 
very sensitive and positionally very accurate; the radio band does not 
suffer from the absorption which maybe a limitation in other wavebands, including the soft X-ray one.
Moreover, the radio properties of a source, such as the spectrum or morphology, could provide further 
insight into their nature. In particular, flat-spectrum radio sources at high galactic 
latitudes are likely to be AGN, possibly associated to a blazar or to the compact core of a radio galaxy.
 
In this work, we present a sample of six unidentified INTEGRAL/IBIS sources, selected on the
basis of their association with a radio source showing a flat spectrum.
To confirm the radio/INTEGRAL association and provide broad-band spectral 
information on these objects, we have also performed X-ray follow-up observations with
the X-ray telescope (XRT; \citealt{burrows:2005}) on board the Swift satellite \citep{gehrels04}. 
The radio and X-rays properties are then combined with information
gathered from the literature to identify the likely counterpart(s) and 
discuss the nature of each INTEGRAL source.

\section{The sample}

The sample selection was made on the basis of the cross-correlation between the fourth
INTEGRAL/IBIS catalogue \citep{Bird:2010} and several public radio catalogues, using the
technique described in \citet{stephen05,stephen06,stephen10} and fully tested in these papers. 
The radio catalogues employed in the cross-correlation analysis are 
from the following surveys: the Combined 
Radio All-sky Targeted Eight GHz Survey (CRATES; \citealt{healey07}), 
the Australia Telescope 20 GHz Survey (AT20G; \citealt{murphy10}), 
the Sydney University Molonglo Sky Survey (SUMSS; \citealt{bock99}), 
and the VLA Low-frequency Sky Survey (VLSS; \citealt{cohen07}).
The cross-correlation procedure returns a list of sources that at least have  
a counterpart in one of those catalogues. Among these sources, we selected 
six INTEGRAL objects that have not yet been identified in other wavebands and have an association with a
flat-spectrum radio source\footnote{A flat spectrum radio source is one with energy 
index $\alpha$$\ge$-0.5, assuming S$_{\nu}$$\propto$$\nu^{\alpha}$}. 
In particular, we find that five out of these six objects are listed in the CRATES catalogue,
which is an all-sky survey of flat-spectrum radio sources.
The remaining object (IGR J18129--0649, see Table~\ref{ibis_log}) does not have a CRATES 
counterpart, but is reported to have a flat radio spectrum in the SPECFIND
catalogue v.2.0 \citep{vollmer09}. It is also
the only object located within ten degrees of the Galactic plane, hence 
the only one for which an extragalactic nature is less secure. 
However, its Galactic latitude is above $|$5$|$ degrees, it is reported as a 
symmetric double in the Texas Survey of Radio sources \citep{douglas96},
and is variable in hard X-rays (see below), all properties that hint toward an 
AGN close to the Galactic plane. For this reason we have retained IGR J18129--0649 
within the set of sources analysed in this work. Finally, we note that  
two sources appear in more than one catalogue:
IGR J14488--4008 in CRATES and SUMSS and IGR J18129--0649 in AT20G and VLSS. 

The sample is presented in Table~\ref{ibis_log}, 
where we list in the first six columns the INTEGRAL name, coordinates, 
90\% positional error radius, bursticity flag if any, 20--40\,keV flux, and 
Galactic column density in the direction of each source \citep{kalberla05}. 
Hard X-ray information is all taken from Bird et al. (2010).
The bursticity of a source is defined as the ratio of the maximum significance 
on any timescale, compared to the significance defined for the whole data set analysed 
(see \citealt{Bird:2010} for details). A bursticity of one defines a persistent source, while a bursticity 
greater than one (flagged Y in the Bird et al. survey) implies that the significance of a source
is increased by the omission of some observations from the analysis, 
presumably when the source was in quiescence. A bursticity greater than four (flagged YY) 
instead indicates a strongly variable source.
We note that all but two objects are classified as variable in the fourth INTEGRAL/IBIS
catalogue; the two exceptions are IGR J11544--7618 and IGR J14488--4008,
which are reported as persistent sources.
In the last three columns of Table~\ref{ibis_log}, we list the coordinates of the 
flat-spectrum radio association found and its energy spectral index.
The radio positions are taken from CRATES, except for IGR J18129--0649,
for which we use the NVSS coordinates \citep{condon98}. The radio spectral 
indices are all from CRATES (estimated between 0.843/1.4 and 4.85\,GHz),
except for IGR J18129--0649, for which the SPECFIND catalogue v.2.0 
has been used instead \citep{vollmer09}.

\begin{table*}
\centering
\footnotesize
\caption{{\bf The Sample: INTEGRAL/IBIS data and radio associations}}
\label{ibis_log}
\begin{tabular}{lcrccccccc}
\hline
\hline
{\bf Name}&{\bf RA}&{\bf DEC}& {\bf Pos. err.}& {\bf Burst.}& {\bf F$_{20-40}^{\dagger}$}&{\bf N$_{\rm H}$}&\multicolumn{2}{c}{\bf Radio Coordinates$^{\dagger}$}
&{\bf $\alpha_{radio}$} \\
          &         &        & {\bf (arcmin)} &             &  &{\bf 10$^{22}$cm$^{-2}$}&{\bf RA (J2000)} &{\bf Dec (J2000)}&  \\ 
\hline
\hline
IGR J06073-0024 & 91.830 & -0.415 & 4.8 & YY &  1.21 (5.98)$^{\ast}$ & 0.23 &06 06 57.44 & -00 24 57.5 & -0.15$^C$ \\
IGR J07225-3810 &110.621 & -38.168 & 5.3 & Y &  0.98 (1.74)$^{\ast}$& 0.21 &07 22 22.00 & -38 14 55.0  & -0.26$^C$ \\
IGR J11544-7618 &178.592 &-76.309 & 5.0 &    &  0.68      & 0.08 &11 55 47.00 & -76 19 08.0  & +0.82$^C$  \\
IGR J14488-4008 &222.209 &-40.142 & 4.4 &    &  0.38      & 0.07 &14 48 51.01 & -40 08 45.7  & -0.03$^C$  \\
IGR J18129-0649 &273.224 & -6.829 & 4.1 & Y  &  0.30 (0.45)$^{\ast}$ & 0.35 &18 12 50.95 & -06 48 23.8$\ddagger$  & -0.49$^S$ \\
IGR J19386-4653 &294.653 &-46.886 & 4.5 & YY & $<$0.30 (5.30)$^{\ast}$     & 0.05 &19 38 26.32 & -46 57 25.6  & +0.26$^C$ \\
\hline
\multicolumn{10}{p{1.\linewidth}}{$^{\dagger}$ Fluxes are expressed in units of 10$^{-11}$erg\,cm$^{-2}$\,s$^{-1}$}\\
\multicolumn{10}{p{1.\linewidth}}{$^{\ast}$ The flux reported in parenthesis refers to the peak flux in 20-40\,keV band for sources
detected through the bursticity analysis (see text for details).}\\

\multicolumn{10}{p{1.\linewidth}}{$^{\ddagger}$ Radio position from the NVSS catalogue (see text) }\\

\multicolumn{10}{p{1.\textwidth}}{{\bf Notes}: energy spectral index  from  C=CRATES (estimated in the range 0.843/1.4  and  4.85 GHz) ; S=SPECFIND v.2.0.}\\
\end{tabular}
\end{table*}

\section{Swift observations and data reduction}

To assess the likelihood of the INTEGRAL/radio association and to 
obtain information over a wide range of frequencies, 
we requested and obtained follow-up observations of all six objects 
with the X-ray telescope (XRT; \citealt{burrows:2005})
on board the Swift satellite \citep{gehrels04}. We note that no X-ray data 
were available for these INTEGRAL sources prior to the XRT pointings.
For each source, one or more measurements were performed as reported in Table~\ref{xrt_log}; 
multiple observations were summed together to enhance the signal-to-noise ratio.
XRT data reduction was done using the XRTDAS standard data pipeline
package ({\sc xrtpipeline} v.0.12.6 included in the HEASOFT package v.6.12)\footnote{The XRT Data
Analysis Software (XRTSDAS) was developed by the ASI Science Data Center (ASDC), Italy 
in collaboration with HEASARC NASA/GSFC and is available at http://heasarc.gsfc.nasa.gov/docs/swiftanalysis/ }
in order to produce screened event files. All data were extracted only in the photon counting (PC) mode 
\citep{Hill:2004}, by adopting the standard grade filtering (0--12 for PC) according to the XRT nomenclature.

The XRT images extracted in the 0.3--10\,keV band were searched for significant excesses (above 2.5$\sigma$
level) falling within the IBIS positional uncertainty 
or close to it. One out of ten counterparts is in fact 
expected to fall just outside the 90$\%$ IBIS error circle.
Further details on the image analysis can be found in \citet{landi10},
while the present results are reported in Table~\ref{xrt_log}, where we list, besides
the INTEGRAL source name and number of X-ray observations performed, the total exposure available, 
the position and relative uncertainty of each X-ray detection and the 
number of sigma found. Sources detected above 3\,keV are highlighted,
since these are the most likely to be true counterparts of the IBIS sources.
Figures~\ref{06073} to \ref{19386} show the collection of NVSS/SUMSS
image cut-outs for all sources, with the IBIS
error circle and the position of the XRT detection(s) superimposed. 
In all but one case, that of IGR J11544--7618, the flat spectrum 
radio association reported in Table~\ref{ibis_log} has been  
detected in X-rays, suggesting that it could be a viable  counterpart. 

We also studied the X-ray spectral properties of each source  in order to gain further information.
Events for spectral analysis were extracted within a circular region of radius 20\arcsec,
centered on the source position, which encloses about 90\% of the PSF at 1.5\,keV (see \citealt{moretti04}).
The background was taken from various source-free regions close to the X-ray source of interest, using
circular regions with different radii in order to ensure an evenly
sampled background. Spectra were extracted from the corresponding event files using the \texttt{XSELECT} 
software and binned using the \texttt{FTOOL} \texttt{grppha} in an appropriate way, so that the $\chi^{2}$ 
statistic could be applied. We used version v.011 of the response matrices and created individual
ancillary response files \textit{arf} using \texttt{xrtmkarf} v. 0.5.9.

Analysis of these spectra was performed in the 0.3-6\,keV energy range, 
using \texttt{XSPEC} version 12.5.1 \citep{Arnaud:1996}. 
The data were fitted using a simple power law absorbed by the Galactic
column density measured in the direction of each source and reported in Table~\ref{ibis_log}. Errors on photon index  
are quoted at 90\% confidence level for one parameter of interest ($\Delta\chi^2$=2.71). 
The results of the spectral analysis are reported in the last two columns of Table~\ref{xrt_log},
where we list the photon index and the estimated 2--10\,keV flux.

\begin{table*}
\centering
\small
\caption{{\bf The Sample: Swift/XRT data}}
\label{xrt_log}
\begin{tabular}{lccclccccc}
\hline
\hline
{\bf Name}&{\bf N. Obs}&{\bf Exposure}&{\bf RA}&\multicolumn{1}{c}{\bf DEC}& {\bf Pos. err.}& {\bf $\sigma$}&{\bf Radio Ass.$^{\star}$}&{\bf $\Gamma$}&{\bf F$_{\bf\rm 2-10keV}$}\\
          &            &  {\bf (sec)} &        &         & {\bf (arcsec)} &               &                      &              &{\bf erg\,cm$^{-2}$\,s$^{-1}$}       \\ 
\hline
\hline

IGR J06073-0024 & 4 &  3961  & 06 06 57.20 & -00 24 54.70       & 6.0 & 3.0             & Y & 1.8f                  & 1.10$\times$10$^{-13}$\\
IGR J07225-3810 & 3 &  4488  & 07 22 27.90 & -38 14 55.39 (N1)  & 4.7 & 7.0$^{\dagger}$ & Y & 1.52$^{+0.45}_{-0.46}$&5.76$\times$10$^{-13}$\\   
                &   &        & 07 21 58.30 & -38 12 44.40 (N2)  & 6.0 & 4.1$^{\dagger}$ & N & 1.8f      &1.1$\times$10$^{-13}$\\     
 
IGR J11544-7618 & 2 & 4186   & 11 53 16.46 & -76 18 54.98       & 5.5 & 4.6             & N & 1.8f                  &6.18$\times$10$^{-14}$\\

IGR J14488-4008 & 3 & 8784   & 14 48 50.97 & -40 08 47.01 (N1)  & 3.8 & 14.1$^{\dagger}$& Y & 2.20$^{+0.68}_{-0.68}$&2.53$\times$10$^{-12}$\\
                &   &        & 14 48 50.82 & -40 10 56.70 (N2)  & 4.6 & 5.8             & N & 1.8f                  &1.42$\times$10$^{-13}$\\ 
IGR J18129-0649 & 1 & 1314   & 18 12 50.68 & -06 48 25.11       & 4.6 & 7.9$^{\dagger}$ & Y&1.44$^{+0.41}_{-0.43}$&2.74$\times$10$^{-12}$ \\
IGR J19386-4653 & 1 & 5114   & 19 38 59.80 & -46 51 24.20 (N1)  &  6  & 3.0             & N & 1.8f                  &8.00$\times$10$^{-14}$\\
                &   &        & 19 38 26.20 & -46 57 20.10 (N2)  &  6  & 2.7             & Y & 1.8f                  &6.00$\times$10$^{-14}$\\
\hline
\multicolumn{10}{p{.8\textwidth}}{$^{\dagger}$: source detected also above 3 keV.}\\
\multicolumn{10}{p{.8\textwidth}}{$^{\star}$: confirmed association between X-ray detection and radio counterpart listed in Table 1.}\\
\end{tabular}
\end{table*}

\section{Source details}

In the following, we examine each individual INTEGRAL source in detail 
to assess the likelihood of the X/radio association and understand the true nature of each object.
To this end, we use all the available multiwavelength information in the literature or 
in publicly available databases; we also
analyse the source radio loudness, defined 
as Log(R$_{\rm L}$) = Log(L$_{\rm 5GHz}$/L$_{\rm B}$)\footnote{Radio loud AGN have Log(R$_{\rm L}$) 
values greater than 1 \citep{Kellermann:1989}}, where L$_{\rm 5GHz}$ is the luminosity
at 5\,GHz and L$_{\rm B}$ the luminosity in the B band. 
We also use the S$_{\rm I}$ statistics developed by \citet{edelson12} 
and based on WISE (Wide-Field Infrared Survey Explorer; \citealt{wright10}) 
and 2MASS (Two Micron All-Sky Survey; \citealt{skrutskie06}) 
data to reliably identify IGR J18129--0649 as an AGN: sources for which S$_{\rm I}$
is below 1.73 have a 95\% likelihood of being an AGN of type 1
(Seyfert1, Quasar or Blazar). WISE colours have also been used
to probe the blazar likelihood of some objects \citep{massaro12}. In fact, in the [3.4]-[4.6]-[12] micron
colour-colour diagram, Blazars cover a distinct region, hereafter the WISE blazar strip 
(WBS; see also Figure 1 in \citealt{massaro12}):
if a source falls within or close to the WBS, it 
can potentially be an AGN of the blazar type.

\subsection{IGR J06073--0024}

\begin{figure}
\centering
\includegraphics[width=0.9\linewidth]{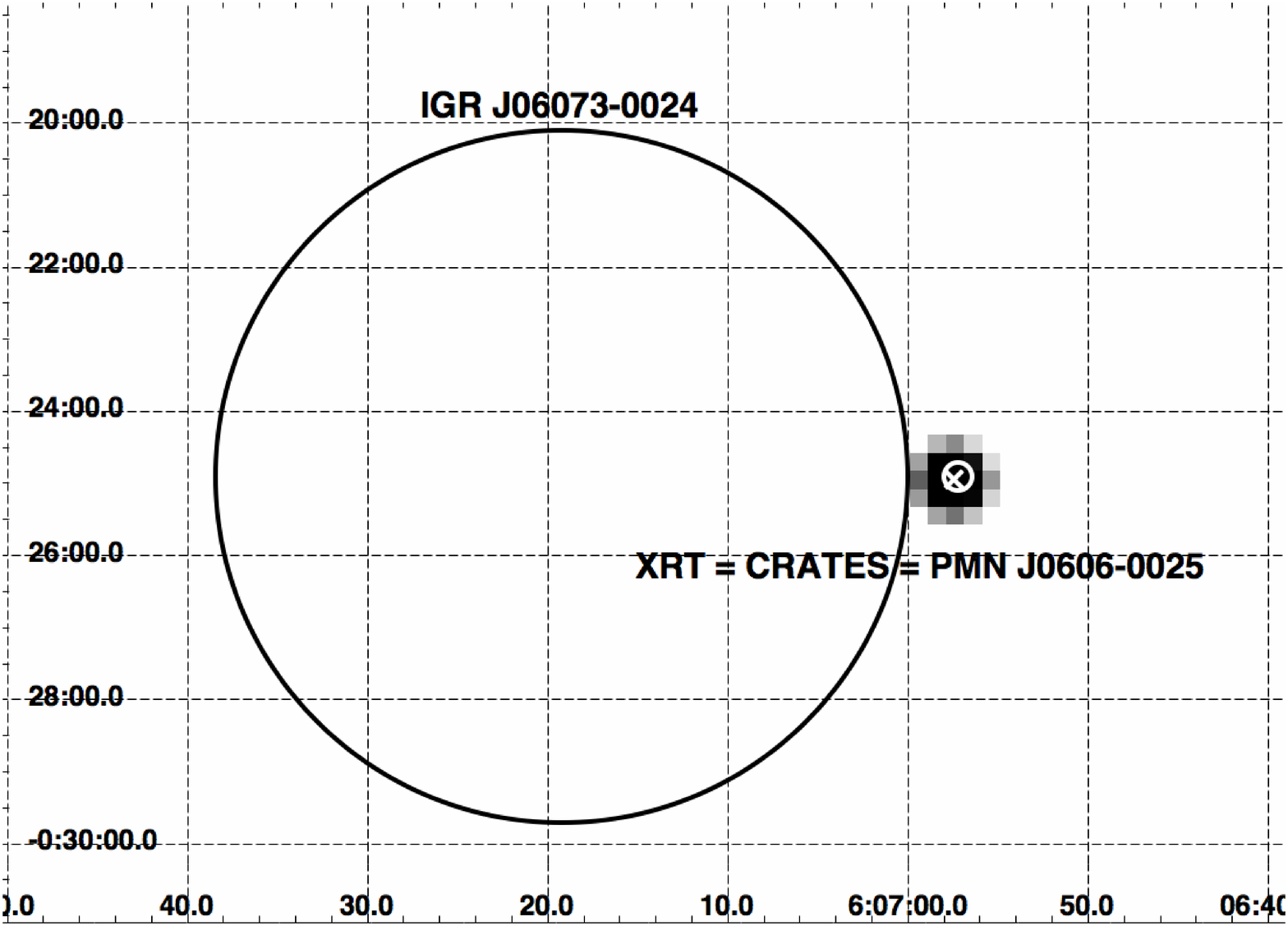}\\
\caption{NVSS 1.4\,GHz  image of the region surrounding IGR J06073--0024.
 The large circle refers to the IBIS positional uncertainty, and the smaller 
one to the XRT position, which is coincident with the CRATES/PMN source  
(indicated with an X-shaped symbol).}
\label{06073}
\end{figure}

Just outside the INTEGRAL 90\% error circle
(see Figure~\ref{06073} and Table~\ref{xrt_log}), we find an X-ray source that is consistent 
with the counterpart found in the CRATES catalogue and reported in NED as PMN J0606-0025. 
It is detected at several radio frequencies from 1.4 to 8.4\,GHz and shows
a compact radio morphology (see Figure~\ref{06073}). It has a NVSS 1.4\,GHz flux of 133.9$\pm$4\,mJy,
it is detected at 5\,GHz in the Parkes-MIT-NRAO (PMN; \citealt{griffith95}) survey
with a flux of 116$\pm$12\,mJy and in the MIT-Green Bank survey \citep{bennett86} 
with a flux of 96\,mJy, therefore some degree of flux variability at this frequency 
cannot be ruled out. In the CRATES catalogue, the source is reported to have an 8.4\,GHz
flux of $\sim$98\,mJy, which combined with the 1.4\,GHz flux, provides the flat 
spectral index reported in Table~\ref{ibis_log}. The 4.85\,GHz flux can be used together 
with the B magnitude to estimate the source radio loudness.
Indeed, coincident with the radio and X-ray positions, we find an optical counterpart in the USNO B-1 
catalogue \citep{monet03}, with magnitude B=20.2 (or 0.02\,mJy in flux); 
here and in the following we use the 
standard photometric system conversion from magnitude to flux \citep{zombeck90}. 
The radio loudness therefore lies in the range 3.7-3.8.

The CRATES source has recently been optically 
classified and found to be a broad emission line AGN (FWHM = 5600\,km\,s$^{-1}$) at z=1.08 
\citep{masetti12}). This initially tentative identification is now supported by the X-ray detection. 
The statistical quality of the X-ray spectrum is such that we can only 
estimate a 2--10\,keV flux of 1.1$\times$10$^{-13}$\,erg\,cm$^2$\,s$^{-1}$. 
At the reported redshift and assuming H$_{\rm 0}$=71\,km\,s$^{-1}$\,Mpc$^{-1}$, 
$\Omega_{\Lambda}$=0.73 and $\Omega_{\rm M}$=0.27, the source X-ray luminosities are   
3.8$\times$10$^{43}$\,erg\,s$^{-1}$ and 4.1$\times$10$^{45}$\,erg\,s$^{-1}$ 
in the 2--10\,keV and 20--40\,keV band, respectively; this suggests that the IGR J06073--0024
X-ray spectrum rises towards higher frequencies, as expected in flat spectrum radio quasars.
Masetti and coworkers also estimate the mass of the black hole at the centre
of this object to be 5$\times$10$^8$M$_{\odot}$, i.e. quite a massive black hole,
similar to those found in other high-redshift AGN discovered in hard X-ray surveys
\citep{ghisellini10, derosa12}.

The source is also reported in the WISE survey \citep{wright10} 
with the following magnitudes W$_1$=14.99 (3.4\,$\mu$m),
W$_2$=14.03 (4.6\,$\mu$m), W$_3$=11.51 (12\,$\mu$m) and W$_4$=8.83 (22\,$\mu$m).
Its WISE colours are W$_2$-W$_3$=2.52 and W$_1$-W$_2$=0.96, therefore placing it 
in the WBS. We therefore conclude that IGR J06073--0024 is a newly discovered 
flat spectrum radio quasar at high z, hence a new hard X-ray selected 
blazar.

\subsection{IGR J07225--3810}
\begin{figure}
\centering
\includegraphics[width=0.9\linewidth]{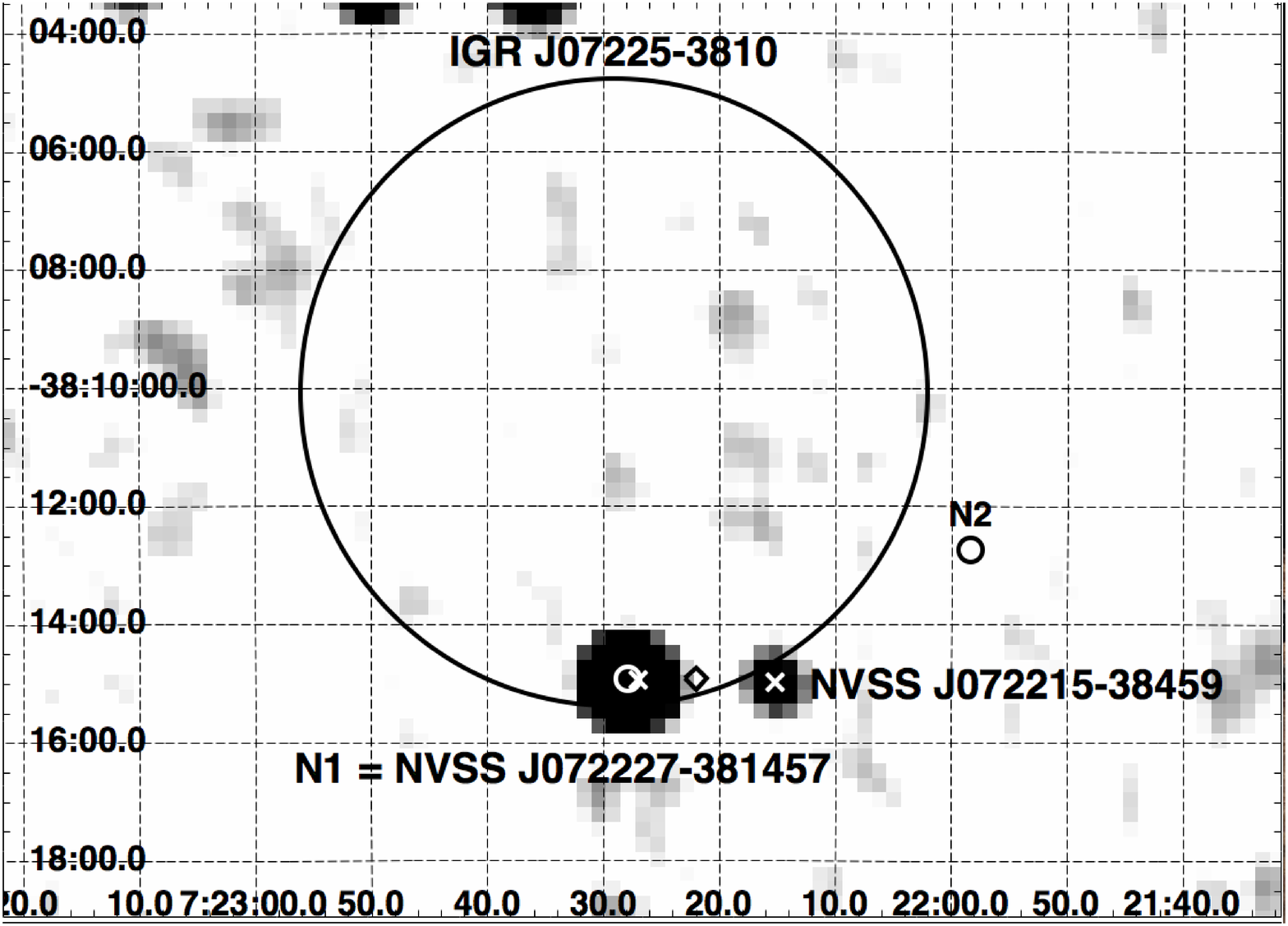}
\caption{NVSS 1.4\,GHz image of the region surrounding IGR J07225--3810. 
The large circle refers to the IBIS positional uncertainty, and the smaller ones to 
the position of the two XRT detections. The X-shaped
symbol corresponds to the position of the radio counterpart. Also shown with a small 
diamond is the position 
of the CRATES/PMN source discussed in the text, which falls in between the two NVSS detections.}
\label{07225}
\end{figure}

In this case, we find two potential X-ray counterparts (see Table~\ref{xrt_log}): 
one (N1 in Figure~\ref{07225} and Table~\ref{xrt_log})
lies inside the IBIS error circle, while the other (N2) is outside its border.
Source N1 is also the brightest of the two and the only one detected above 3\,keV 
with some confidence (above 2.5 sigma); this, together with its 
location inside the INTEGRAL positional uncertainty, makes it 
a more convincing counterpart to the INTEGRAL source. 
Besides, we could not find any obvious association for source N2, so  
this object cannot be discussed further. 

Source N1 coincides instead with the CRATES association proposed in 
Table~\ref{ibis_log} and reported in NED as PMN J0722-3814.
As is evident in Figure~\ref{07225}, this CRATES source is slightly offset
from  the bright radio source (NVSS J072227-381457) falling inside the 
INTEGRAL error circle. This offset may be due to the
inability of the PMN survey (used as a reference for the CRATES position) 
to resolve the emission in two nearby objects as done by the NVSS
(see Figure~\ref{07225}). In the following, we assume that the bright NVSS source and 
the CRATES/PMN object are the same: in this case the 1.4\,GHz (NVSS) and 0.843\,GHz
(SUMSS) fluxes are 135 and 203\,mJy respectively, while the flux at 4.85\,GHz (PMN) 
is $\le$ 110\,mJy (PMN) (due to possible contamination from the nearby 
and dim NVSS detection). The 1 to 5\,GHz spectral index is around -0.2/-03, 
i.e. compatible with that reported in Table~\ref{ibis_log}. 
Consistent with the NVSS/XRT positions, we find an optical USNO B-1 source
with optical magnitude B=19.1 (0.1\,mJy in flux), which provides Log(R$_{\rm L}$)$\le$ 3.
 
The XRT spectrum for source N1 has enough statistics to allow determination of the 
photon index, which, despite the large errors, is rather flat, at 1.52$^{+0.45}_{-0.46}$;
the X-ray 2--10\,keV flux is 5.76$\times$10$^{-13}$\,erg\,cm$^2$\,s$^{-1}$. 
The source is also detected by  WISE with the following  magnitudes: W$_1$=14.67,
W$_2$=13.61, W$_3$=10.22 and W$_4$=8.33. Its WISE colours, 
W$_2$-W$_3$=3.39, and W$_1$-W$_2$=1.06, locate this source on the WISE blazar
strip in a region compatible with the colours of flat spectrum radio quasars.
We therefore conclude that source N1 is an AGN possibly of the blazar 
type given its overall properties (flat X-ray spectrum, WISE colours,
and radio characteristics); as such, it is a convincing counterpart of 
the variable INTEGRAL source. The nature of object N2 is 
unclear  at this stage, but its location, as well as X-ray 
weakness, makes it a less convincing association.

\subsection{IGR J11544--7618}

\begin{figure}
\centering
\includegraphics[width=0.85\linewidth]{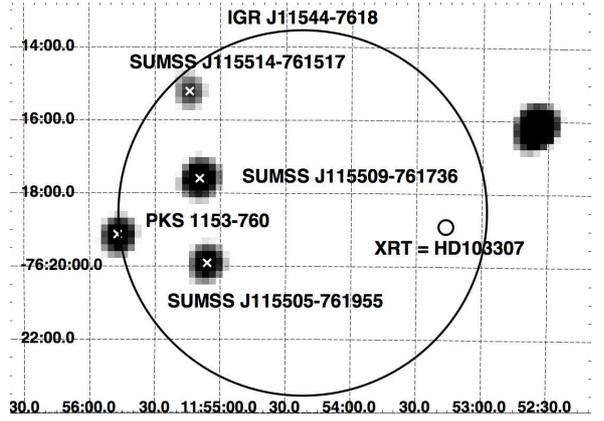}\\
\caption{SUMSS 0.84\,GHz  image of the region surrounding IGR J11544--7618. 
The large circle refers to the IBIS positional uncertainty, 
and the smaller one to the XRT position of the star HD103307. 
Also highlighted are the position of the CRATES source (PKS 1153-780) 
and of 3 SUMSS detections all marked by X-shaped symbols.}
\label{11544}
\end{figure}

The region surrounding IGR J11544--7618 is quite complex at radio frequencies. 
Apart from the CRATES source reported in Table~\ref{ibis_log} and listed in NED as PKS 1153--783,
there are three other objects from the SUMSS survey falling within the 
INTEGRAL error circle (see Figure~\ref{11544}). None of these objects is detected 
in X-rays by XRT, placing a limit on their X-ray emission at around 
3$\times$10$^{-14}$\,erg\,cm$^2$\,s$^{-1}$.
 
However, inside the IBIS error circle there is an X-ray source that is not coincident with
any radio emission (see Figure~\ref{11544}) and previously reported as a
Rosat faint source (1RXS J115313.6-761935\footnote{The Rosat Faint Catalogue is 
available at http://www.xray.mpe.mpg.de/rosat/survey/rass-fsc/}); 
the X-ray spectrum from this source is soft, since it is not detected above 3\,keV.
The XRT coordinates coincide with a bright star, HD103307 of spectral type  G0/3IV/V. 
Solar-like stars like HD103307 emit X-rays via coronal emission, but are unlikely
to be detected at energies higher than a few keV.
However, they sometimes show activity in the form of energetic flares, and 
these stellar flares radiate at all wavelengths from radio to gamma-rays, 
last typically from minutes to hours/days, and have been 
occasionally detected by IBIS as variable objects. 
However, IGR J11544--7618 is a persistent source in the 20--40\,keV band,
so we can exclude that the emission detected by INTEGRAL is due to flares 
coming from this star. 

At the same time we cannot invoke X-ray variability
to explain the lack of X-ray emission from the CRATES or any 
other radio source in the region. 
Unfortunately, the counterpart of this INTEGRAL object remains unclear, 
and only further follow-up observations in various wavebands would 
be able to shed some light on its nature.
Another possibility that cannot be ruled out at the moment is that
IGR J11544--7618 is a spurious detection by IBIS.

\subsection{IGR J14488--4008}

\begin{figure}
\centering
\includegraphics[width=0.85\linewidth]{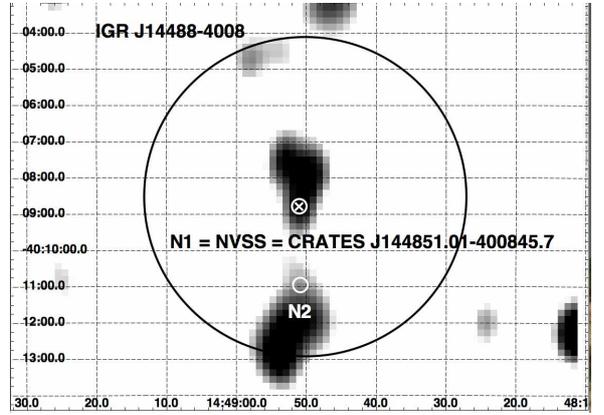}
\caption{SUMSS 0.84\,GHz  image of the region surrounding IGR J14488--4008. 
The large circle refers to the IBIS positional 
uncertainty, and the smaller ones to the 
the position of the two XRT detections. The X-ray source N1 coincides with one of the 
radio excesses associated to  
CRATES source (indicated with an X-shaped symbol) and likely representing the compact core of a radio galaxy.}
\label{14488}
\end{figure}

\begin{figure}
\centering
\includegraphics[width=0.85\linewidth]{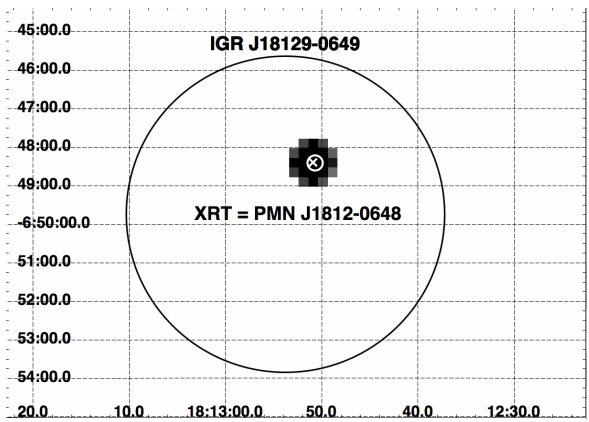}\\
\caption{NVSS 1.4\,GHz image of the region surrounding IGR J18129--0649. 
The large circle refers to the IBIS positional uncertainty, and the smaller 
one to the XRT position, which is coincident with a radio source PMN J1812-0648
(indicated with an X-shaped symbol) detected in various catalogues.}
\label{18129}
\end{figure}

\begin{figure}
\centering
\includegraphics[width=0.85\linewidth]{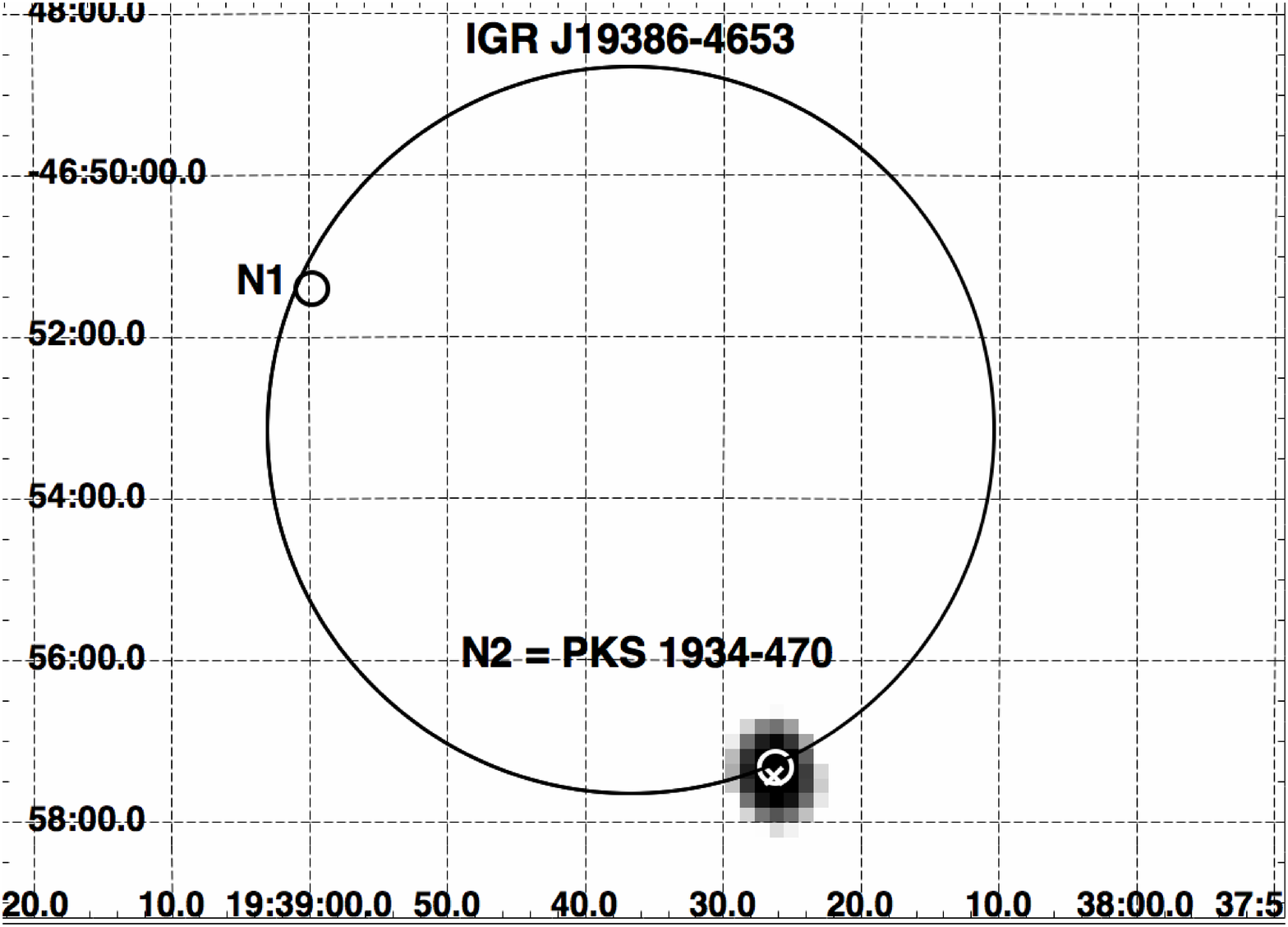}
\caption{SUMSS 0.8\,GHz image of the region surrounding  IGR J19386--4653. 
The large circle refers to the IBIS positional 
uncertainty, and the smaller ones to the 
the position of the two XRT detections. The X-ray source N1 is not detected 
in radio, while source N2 is coincident with the CRATES/PKS source (indicated with an X-shaped symbol).}
\label{19386}
\end{figure}

As is evident from Figure~\ref{14488}, the CRATES source listed in Table~\ref{ibis_log}
has an extremely complex radio morphology, with multiple structures that resemble 
a radio galaxy with two bright lobes fully contained within the INTEGRAL error circle.
Also in this case, there are two X-ray sources (N1 and N2 in Table~\ref{xrt_log}
and in Figure~\ref{14488}) falling  within IBIS positional uncertainty; 
one of the two (N1) is brighter and harder since it is also detected above 3\,keV.
This source is associated to one of the CRATES excesses (CRATES J1448-4008), 
as well as to one of the NVSS (NVSS J144851-400846) peaks reported in this region. 
This excess is located at the centre of 
the radio structure and is also coincident with the optical position of the 
main galaxy, PGC589690 \citep{paturel03}.
N1 could therefore be  the compact core of a newly discovered radio galaxy.

Because of the complex structure, it is difficult to combine radio observations 
at various frequencies to estimate a spectral index.
If we confine the search to radio emission within the X-ray error circle of source N1, 
we find that it is detected at 1.4, 4.85, 8.4, and 8.6\,GHz 
with fluxes of 48, 48, 45.9, and 46\,mJy respectively (NVSS; CRATES; ATPMN; \citealt{mcconnell12}). 
This provides a spectral index similar to the one reported in Table~\ref{ibis_log}. 
In the optical band, we find a counterpart in the USNO-B1 catalogue having B$\sim$14.2 (or 8.9 mJy in flux); 
the radio loudness parameter is therefore 0.73, making this object radio quiet.

The X-ray data are well-fitted using a double power-law
model with indices tight together and with $\Gamma$ = 2.20, consistent with what
reported by \citet{malizia11}. In addition to the Galactic absorption, the  
data require an intrinsic column density N$_{\rm H}$ = 
(6.2$^{+2.1}_{-1.9}$)$\times$10$^{22}$\,cm$^{-2}$, suggesting that it might be a 
type 2 AGN.

Source N2 cannot be characterised with the same detail. In X-rays, it
is much dimmer and softer than object N1 and is also 
variable, since seen in only one of the three XRT pointings, which contrasts 
with the persistent nature of IGR J14488--4008. 
It has a counterpart in the USNO-B1/2MASS catalogues, but no detection
in radio. It is difficult to assess its nature at this stage, but its X-ray characteristics 
(flux variability, X-ray weakness, as well as X-ray spectral softness) suggest 
that this source is unlikely to be the counterpart of the persistent 
INTEGRAL object, thus making N1 the most likely association to IGR J14488--4008.

\subsection{IGR J18129--0649}

\begin{figure}
\centering
\includegraphics[width=0.85\linewidth]{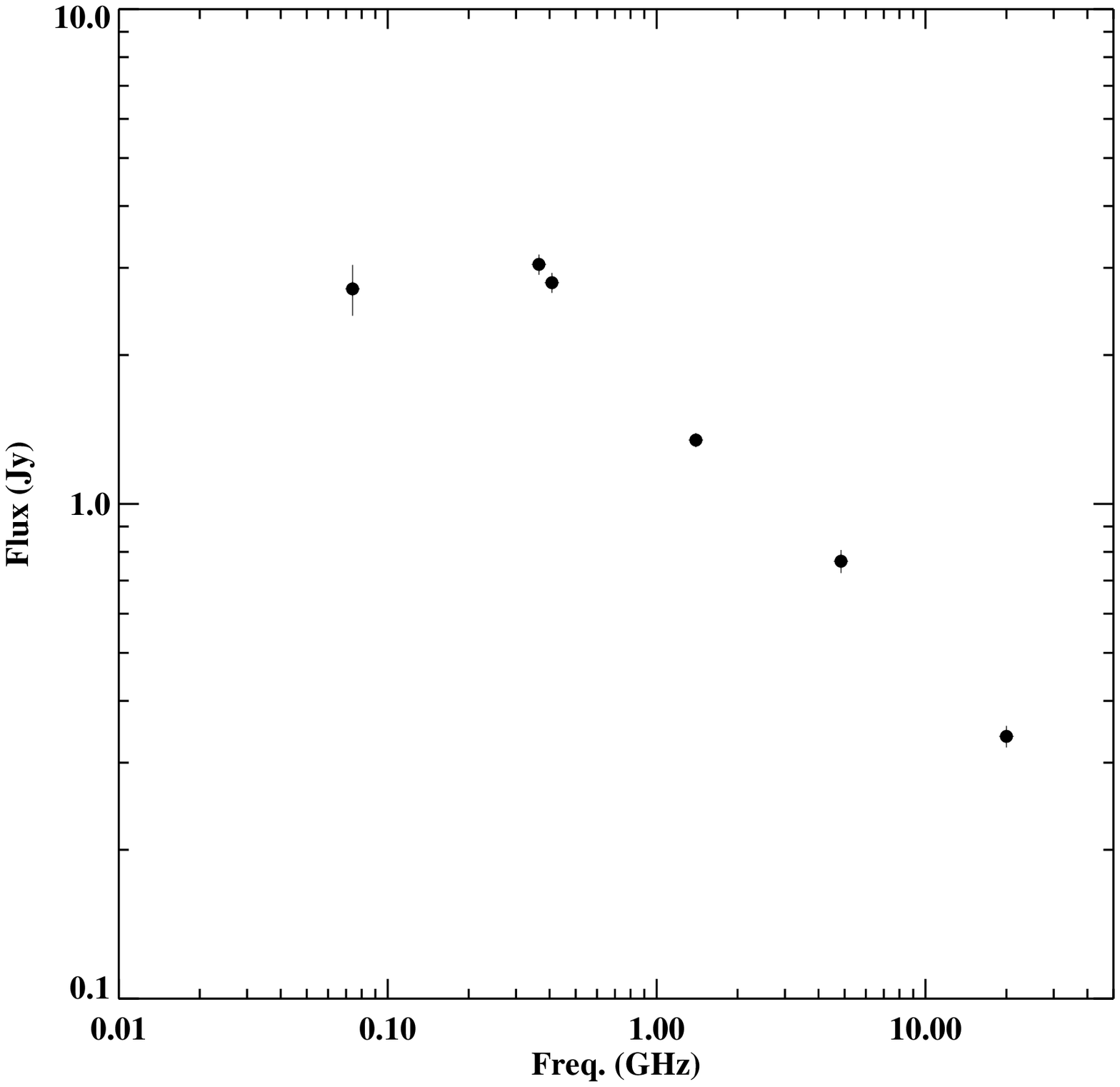}\\
\caption{Spectrum of the radio source PMN J1812-0648 associated to IGR J18129--0649. Data are from VLSS \citep{cohen07},
Texas \citep{douglas96}, Molonglo \citep{bock99}, NVSS \citep{condon92}
PMN \citep{griffith95}, and AT20G \citep{murphy10} surveys.}
\label{18129_spe}
\end{figure}

The radio source listed in Table~\ref{ibis_log} 
lies well within the INTEGRAL error box.
It coincides with the only detection made by XRT in this region 
(see Table~\ref{xrt_log} for details), which suggests that it is a valid 
counterpart of the INTEGRAL source. It is quite bright at radio frequencies, 
with a flux ranging from 2720\,mJy at 0.074\,GHz \citep{cohen07} to 339\,mJy at 20\,GHz \citep{murphy10}.
The overall spectrum is shown in Figure~\ref{18129_spe}: 
it has an energy index $\alpha$$\sim$-0.5 
from 0.365 to 20\,GHz with some deviation at low frequencies. 
The source, named PMN J1812--0648 in NED, is compact in the NVSS map, 
thus suggesting that the classification as a symmetric double in the Texas survey is 
probably incorrect.

Besides being the brightest radio source among our sample, 
it is also the strongest in X-rays with a 2--10\,keV flux of 
2.7$\times$10$^{-12}$\,erg\,cm$^2$\,s$^{-1}$.
The X-ray spectrum is well-fitted by a simple power-law 
model with a flat photon index of 1.4.

As anticipated, this is the only object in the sample that is relatively close 
to the Galactic plane, hence the only one for which an
AGN nature is uncertain. For this reason we used the method 
employed by \citet{edelson12} to recognise AGN using simply  
WISE/2MASS colours. The source is listed in the WISE catalogue with magnitudes
W$_1$=13.13, W$_2$=11.80, W$_3$=9.19, and W$_4$=6.77, and the 2MASS magnitudes are 
instead J=16.72, H=15.19 and K=14.77. Using the \citet{edelson12} formalism, 
we estimate S$_{\rm I}$=0.9, i.e. a value  below the cut of 
1.73 used to reliably identified type 1 AGN.
Furthermore, we note that the WISE colours are W$_2$-W$_3$=2.6 and W$_1$-W$_2$=1.34, 
placing the source outside the WiSE blazar strip in a region 
compatible with the location of quasars.

The source has no counterpart in the USNO-B1 catalogue, with only an upper limit
to the B magnitude of 20.7 (or 0.02\,mJy in flux). 
Combining the 4.85\,GHz flux of 766\,mJy measured by the PMN survey
with that in the B band gives a radio loudness parameter Log(R$_{\rm L}$)$\geq$4.5; 
i.e., the source is certainly radio-loud. While the association of PMN J1812-0648 
with the INTEGRAL source is secure and its AGN nature validated by 
observational evidence, its optical class is unclear, 
and only optical follow-up observations can assess its true nature.

\subsection{IGR J19386--4653}

Inside the 90\% IBIS positional uncertainty, XRT detects two weak 
sources (N1 and N2 in Table~\ref{xrt_log} and in Figure~\ref{19386}), 
none of which is visible above 3\,keV. They are both quite dim in X-rays,
with 2--10\,keV fluxes below 10$^{-13}$\,erg\,cm$^2$\,s$^{-1}$. 
We remind the reader that IGR J19386--4653 is extremely variable in INTEGRAL,
which could explain the weakness of both X-ray detections.
The first source has two counterparts in the USNO-B1 and one in the 2MASS catalogues,
but it is not detected in radio, nor it is listed in WISE. 

The second one, N2, is instead associated with the CRATES object (also PKS 1934--470 in NED) 
listed in Table~\ref{ibis_log}, thus suggesting that 
this is a viable counterpart to the INTEGRAL detection.
Source N2 has been previously seen in X-rays since it is listed in 
the WGACAT catalogue \citep{white00}, with a large positional uncertainty 
(50\arcsec) that is, however, compatible with the XRT location. 
The WGACAT source is also reported in the Roma-Blazar catalogue \citep{massaro09} 
and is classified by \citet{landt04} as a flat spectrum radio quasar at z=0.8; unfortunately, 
this source is located 35\arcsec from the XRT detection casting some doubts on their association.
On the other hand, we could not find any optical counterpart within 
the XRT positional uncertainty, thus making the case even more puzzling.

At radio frequencies, source N2/PKS 1934--470 is well studied, with detections 
reported at various frequencies going from 138\,mJy at 0.843\,GHz \citep{bock99}  
to 90\,mJy at 8.6\,GHz \citep{mcconnell12}; this provides a radio spectral 
index of -0.2, still flat but substantially different from 
what is reported in Table~\ref{ibis_log}. We note however 
that the value listed in the table is the one estimated in the 0.843 and 4.85\,GHZ band;
when a wider frequency range is considered the index becomes -0.22 CRATES ($\alpha_2$ in CRATES). 
The radio morphology is that of a compact source.
At 5\,GHz the reported fluxes range from 104\,mJy \citep{mcconnell12} to 
174\,mJy \cite{wright94}. Combining these values with the sensitivity limit of 
the USNO-B1 survey in this region (B$\ge$20.8 or 0.02 mJy in flux), we estimate a radio loudness  
Log(R$_{\rm L}$)$\geq$3.7, which suggests that
N2/PKS 1934--470 is a radio loud AGN.
The WISE magnitudes are W$_1$=15.46, W$_2$=14.92, W$_3$=12.45, and W$_4$=8.85 which  
constrains the WISE colours to be W$_2$-W$_3$=2.45 and W$_1$-W$_2$=0.55. 
This locates the source just outside the WBS.

In conclusion, we find that two sources can be responsible for 
the hard X-ray emission seen by INTEGRAL. Source N2 is very likely an AGN of 
still uncertain class, hence a valid counterpart, while less certain 
is the nature of source N1, which therefore remains a more dubious association.

\section{Conclusions}

We have used a well-tested cross-correlation technique 
to extract a sample of six unidentified/unclassified INTEGRAL
objects that have a flat spectrum  radio counterpart. 
This finding, together with the location above the galactic plane, suggests that 
they may all be AGN and therefore a valid association to the hard X-ray emitters.
To confirm these associations and to study the X-ray properties of these objects,
we made use of X-ray data collected by Swift/XRT through target of opportunity
observations. Each source was analysed in detail using also archival multiwaveband data 
and information gathered from the literature. 

We find that five of the six radio associations proposed in this 
work are also detected in X-rays; furthermore, in three cases they are the only 
counterpart found for the INTEGRAL source. More specifically,
IGR J06073--0024 is a flat-spectrum radio quasar at high redshift,  
IGR J14488--4008 is a new radio galaxy with a complex morphology and a likely type 2 AGN nature,
while  IGR J18129--0649 is an AGN of still uncertain optical classification.
The nature of IGR J07225--3810 and IGR J19386--4653 is  defined less 
well, since we find in both cases another X-ray source in the INTEGRAL error circle;
nevertheless the flat spectrum radio source, probably a radio-loud AGN,  
remains a viable and, in fact, more convincing association in both cases.
Only for  IGR J11544--7618 have we not been able to find a likely counterpart, 
since the radio association is not an X-ray emitter, while the only
X-ray source seen in the field is a G star and therefore unlikely 
to produce persistent hard X-rays as observed by INTEGRAL.



\begin{acknowledgements}
This research has made use of data obtained from the SIMBAD database operated at the CDS, 
Strasbourg, France and from the High Energy Astrophysics Science Archive Research Center (HEASARC), 
provided by NASA$^{\prime}$s Goddard Space Flight Center NASA/IPAC Extragalactic Database (NED).
The authors acknowledge financial support from the ASI under contracts ASI/033/10/0 and ASI/INAF I/009/10/0
\end{acknowledgements}

\bibliography{./mol_biblio}

\begin{thebibliography}{40}
\expandafter\ifx\csname natexlab\endcsname\relax\def\natexlab#1{#1}\fi

\bibitem[{{Arnaud}(1996)}]{Arnaud:1996}
{Arnaud}, K.~A. 1996, in Astronomical Society of the Pacific Conference Series,
  Vol. 101, Astronomical Data Analysis Software and Systems V, ed.
  {G.~H.~Jacoby \& J.~Barnes}, 17

\bibitem[{{Bennett} {et~al.}(1986){Bennett}, {Lawrence}, {Burke}, {Hewitt}, \&
  {Mahoney}}]{bennett86}
{Bennett}, C.~L., {Lawrence}, C.~R., {Burke}, B.~F., {Hewitt}, J.~N., \&
  {Mahoney}, J. 1986, \apjs, 61, 1

\bibitem[{{Bird} {et~al.}(2010){Bird}, {Bazzano}, {Bassani}, {Capitanio},
  {Fiocchi}, {Hill}, {Malizia}, {McBride}, {Scaringi}, {Sguera}, {Stephen},
  {Ubertini}, \& {Dean}}]{Bird:2010}
{Bird}, A.~J., {Bazzano}, A., {Bassani}, L., {et~al.} 2010, \apjs, 186, 1

\bibitem[{{Bock} {et~al.}(1999){Bock}, {Large}, \& {Sadler}}]{bock99}
{Bock}, D.~C.-J., {Large}, M.~I., \& {Sadler}, E.~M. 1999, \aj, 117, 1578

\bibitem[{{Burrows} {et~al.}(2005){Burrows}, {Hill}, {Nousek}, {Kennea},
  {Wells}, {Osborne}, {Abbey}, {Beardmore}, {Mukerjee}, {Short}, {Chincarini},
  {Campana}, {Citterio}, {Moretti}, {Pagani}, {Tagliaferri}, {Giommi},
  {Capalbi}, {Tamburelli}, {Angelini}, {Cusumano}, {Br{\"a}uninger}, {Burkert},
  \& {Hartner}}]{burrows:2005}
{Burrows}, D.~N., {Hill}, J.~E., {Nousek}, J.~A., {et~al.} 2005, \ssr, 120, 165

\bibitem[{{Cohen} {et~al.}(2007){Cohen}, {Lane}, {Cotton}, {Kassim}, {Lazio},
  {Perley}, {Condon}, \& {Erickson}}]{cohen07}
{Cohen}, A.~S., {Lane}, W.~M., {Cotton}, W.~D., {et~al.} 2007, \aj, 134, 1245

\bibitem[{{Condon}(1992)}]{condon92}
{Condon}, J.~J. 1992, \araa, 30, 575

\bibitem[{{Condon} {et~al.}(1998){Condon}, {Cotton}, {Greisen}, {Yin},
  {Perley}, {Taylor}, \& {Broderick}}]{condon98}
{Condon}, J.~J., {Cotton}, W.~D., {Greisen}, E.~W., {et~al.} 1998, \aj, 115,
  1693

\bibitem[{{De Rosa} {et~al.}(2012){De Rosa}, {Panessa}, {Bassani}, {Bazzano},
  {Bird}, {Landi}, {Malizia}, {Molina}, \& {Ubertini}}]{derosa12}
{De Rosa}, A., {Panessa}, F., {Bassani}, L., {et~al.} 2012, \mnras, 420, 2087

\bibitem[{{Douglas} {et~al.}(1996){Douglas}, {Bash}, {Bozyan}, {Torrence}, \&
  {Wolfe}}]{douglas96}
{Douglas}, J.~N., {Bash}, F.~N., {Bozyan}, F.~A., {Torrence}, G.~W., \&
  {Wolfe}, C. 1996, \aj, 111, 1945

\bibitem[{{Edelson} \& {Malkan}(2012)}]{edelson12}
{Edelson}, R. \& {Malkan}, M. 2012, \apj, 751, 52

\bibitem[{{Gehrels} {et~al.}(2004){Gehrels}, {Chincarini}, {Giommi}, {Mason},
  {Nousek}, {Wells}, {White}, {Barthelmy}, {Burrows}, {Cominsky}, {Hurley},
  {Marshall}, {M{\'e}sz{\'a}ros}, {Roming}, {Angelini}, {Barbier}, {Belloni},
  {Campana}, {Caraveo}, {Chester}, {Citterio}, {Cline}, {Cropper}, {Cummings},
  {Dean}, {Feigelson}, {Fenimore}, {Frail}, {Fruchter}, {Garmire}, {Gendreau},
  {Ghisellini}, {Greiner}, {Hill}, {Hunsberger}, {Krimm}, {Kulkarni}, {Kumar},
  {Lebrun}, {Lloyd-Ronning}, {Markwardt}, {Mattson}, {Mushotzky}, {Norris},
  {Osborne}, {Paczynski}, {Palmer}, {Park}, {Parsons}, {Paul}, {Rees},
  {Reynolds}, {Rhoads}, {Sasseen}, {Schaefer}, {Short}, {Smale}, {Smith},
  {Stella}, {Tagliaferri}, {Takahashi}, {Tashiro}, {Townsley}, {Tueller},
  {Turner}, {Vietri}, {Voges}, {Ward}, {Willingale}, {Zerbi}, \&
  {Zhang}}]{gehrels04}
{Gehrels}, N., {Chincarini}, G., {Giommi}, P., {et~al.} 2004, \apj, 611, 1005

\bibitem[{{Ghisellini} {et~al.}(2010){Ghisellini}, {Tavecchio}, {Foschini},
  {Ghirlanda}, {Maraschi}, \& {Celotti}}]{ghisellini10}
{Ghisellini}, G., {Tavecchio}, F., {Foschini}, L., {et~al.} 2010, \mnras, 402,
  497

\bibitem[{{Griffith} {et~al.}(1995){Griffith}, {Wright}, {Burke}, \&
  {Ekers}}]{griffith95}
{Griffith}, M.~R., {Wright}, A.~E., {Burke}, B.~F., \& {Ekers}, R.~D. 1995,
  \apjs, 97, 347

\bibitem[{{Healey} {et~al.}(2007){Healey}, {Romani}, {Taylor}, {Sadler},
  {Ricci}, {Murphy}, {Ulvestad}, \& {Winn}}]{healey07}
{Healey}, S.~E., {Romani}, R.~W., {Taylor}, G.~B., {et~al.} 2007, \apjs, 171,
  61

\bibitem[{{Hill} {et~al.}(2004){Hill}, {Burrows}, {Nousek}, {Abbey}, {Ambrosi},
  {Br{\"a}uninger}, {Burkert}, {Campana}, {Cheruvu}, {Cusumano}, {Freyberg},
  {Hartner}, {Klar}, {Mangels}, {Moretti}, {Mori}, {Morris}, {Short},
  {Tagliaferri}, {Watson}, {Wood}, \& {Wells}}]{Hill:2004}
{Hill}, J.~E., {Burrows}, D.~N., {Nousek}, J.~A., {et~al.} 2004, in Presented
  at the Society of Photo-Optical Instrumentation Engineers (SPIE) Conference,
  Vol. 5165, Society of Photo-Optical Instrumentation Engineers (SPIE)
  Conference Series, ed. {K.~A.~Flanagan \& O.~H.~W.~Siegmund}, 217--231

\bibitem[{{Kalberla} {et~al.}(2005){Kalberla}, {Burton}, {Hartmann}, {Arnal},
  {Bajaja}, {Morras}, \& {P{\"o}ppel}}]{kalberla05}
{Kalberla}, P.~M.~W., {Burton}, W.~B., {Hartmann}, D., {et~al.} 2005, \aap,
  440, 775

\bibitem[{{Kellermann} {et~al.}(1989){Kellermann}, {Sramek}, {Schmidt},
  {Shaffer}, \& {Green}}]{Kellermann:1989}
{Kellermann}, K.~I., {Sramek}, R., {Schmidt}, M., {Shaffer}, D.~B., \& {Green},
  R. 1989, \aj, 98, 1195

\bibitem[{{Landi} {et~al.}(2010){Landi}, {Bassani}, {Malizia}, {Stephen},
  {Bazzano}, {Fiocchi}, \& {Bird}}]{landi10}
{Landi}, R., {Bassani}, L., {Malizia}, A., {et~al.} 2010, \mnras, 403, 945

\bibitem[{{Landt} {et~al.}(2004){Landt}, {Padovani}, {Perlman}, \&
  {Giommi}}]{landt04}
{Landt}, H., {Padovani}, P., {Perlman}, E.~S., \& {Giommi}, P. 2004, \mnras,
  351, 83

\bibitem[{{Malizia} {et~al.}(2011){Malizia}, {Landi}, {Bassani}, {Bird},
  {Gehrels}, \& {Kennea}}]{malizia11}
{Malizia}, A., {Landi}, R., {Bassani}, L., {et~al.} 2011, The Astronomer's
  Telegram, 3290, 1

\bibitem[{{Masetti} {et~al.}(2012){Masetti}, {Parisi},
  {Jim{\'e}nez-Bail{\'o}n}, {Palazzi}, {Chavushyan}, {Bassani}, {Bazzano},
  {Bird}, {Dean}, {Galaz}, {Landi}, {Malizia}, {Minniti}, {Morelli},
  {Schiavone}, {Stephen}, \& {Ubertini}}]{masetti12}
{Masetti}, N., {Parisi}, P., {Jim{\'e}nez-Bail{\'o}n}, E., {et~al.} 2012, \aap,
  538, A123

\bibitem[{{Massaro} {et~al.}(2009){Massaro}, {Giommi}, {Leto}, {Marchegiani},
  {Maselli}, {Perri}, {Piranomonte}, \& {Sclavi}}]{massaro09}
{Massaro}, E., {Giommi}, P., {Leto}, C., {et~al.} 2009, VizieR Online Data
  Catalog, 349, 50691

\bibitem[{{Massaro} {et~al.}(2012){Massaro}, {D'Abrusco}, {Tosti}, {Ajello}, \&
  {Gasparrini}}]{massaro12}
{Massaro}, F., {D'Abrusco}, R., {Tosti}, G., {Ajello}, M., \& {Gasparrini},
  A.~P.~D. 2012, ArXiv e-prints

\bibitem[{{McConnell} {et~al.}(2012){McConnell}, {Sadler}, {Murphy}, \&
  {Ekers}}]{mcconnell12}
{McConnell}, D., {Sadler}, E.~M., {Murphy}, T., \& {Ekers}, R.~D. 2012, \mnras,
  422, 1527

\bibitem[{{Monet} {et~al.}(2003){Monet}, {Levine}, {Canzian}, {Ables}, {Bird},
  {Dahn}, {Guetter}, {Harris}, {Henden}, {Leggett}, {Levison}, {Luginbuhl},
  {Martini}, {Monet}, {Munn}, {Pier}, {Rhodes}, {Riepe}, {Sell}, {Stone},
  {Vrba}, {Walker}, {Westerhout}, {Brucato}, {Reid}, {Schoening}, {Hartley},
  {Read}, \& {Tritton}}]{monet03}
{Monet}, D.~G., {Levine}, S.~E., {Canzian}, B., {et~al.} 2003, \aj, 125, 984

\bibitem[{{Moretti} {et~al.}(2004){Moretti}, {Campana}, {Tagliaferri}, {Abbey},
  {Ambrosi}, {Angelini}, {Beardmore}, {Br{\"a}uninger}, {Burkert}, {Burrows},
  {Capalbi}, {Chincarini}, {Citterio}, {Cusumano}, {Freyberg}, {Giommi},
  {Hartner}, {Hill}, {Mori}, {Morris}, {Mukerjee}, {Nousek}, {Osborne},
  {Short}, {Tamburelli}, {Watson}, \& {Wells}}]{moretti04}
{Moretti}, A., {Campana}, S., {Tagliaferri}, G., {et~al.} 2004, in Society of
  Photo-Optical Instrumentation Engineers (SPIE) Conference Series, Vol. 5165,
  Society of Photo-Optical Instrumentation Engineers (SPIE) Conference Series,
  ed. {K.~A.~Flanagan \&amp; O.~H.~W.~Siegmund}, 232--240

\bibitem[{{Murphy} {et~al.}(2010){Murphy}, {Sadler}, {Ekers}, {Massardi},
  {Hancock}, {Mahony}, {Ricci}, {Burke-Spolaor}, {Calabretta}, {Chhetri}, {de
  Zotti}, {Edwards}, {Ekers}, {Jackson}, {Kesteven}, {Lindley}, {Newton-McGee},
  {Phillips}, {Roberts}, {Sault}, {Staveley-Smith}, {Subrahmanyan}, {Walker},
  \& {Wilson}}]{murphy10}
{Murphy}, T., {Sadler}, E.~M., {Ekers}, R.~D., {et~al.} 2010, \mnras, 402, 2403

\bibitem[{{Paturel} {et~al.}(2003){Paturel}, {Petit}, {Prugniel}, {Theureau},
  {Rousseau}, {Brouty}, {Dubois}, \& {Cambr{\'e}sy}}]{paturel03}
{Paturel}, G., {Petit}, C., {Prugniel}, P., {et~al.} 2003, \aap, 412, 45

\bibitem[{{Skrutskie} {et~al.}(2006){Skrutskie}, {Cutri}, {Stiening},
  {Weinberg}, {Schneider}, {Carpenter}, {Beichman}, {Capps}, {Chester},
  {Elias}, {Huchra}, {Liebert}, {Lonsdale}, {Monet}, {Price}, {Seitzer},
  {Jarrett}, {Kirkpatrick}, {Gizis}, {Howard}, {Evans}, {Fowler}, {Fullmer},
  {Hurt}, {Light}, {Kopan}, {Marsh}, {McCallon}, {Tam}, {Van Dyk}, \&
  {Wheelock}}]{skrutskie06}
{Skrutskie}, M.~F., {Cutri}, R.~M., {Stiening}, R., {et~al.} 2006, \aj, 131,
  1163

\bibitem[{{Stephen} {et~al.}(2010){Stephen}, {Bassani}, {Landi}, {Malizia},
  {Sguera}, {Bazzano}, \& {Masetti}}]{stephen10}
{Stephen}, J.~B., {Bassani}, L., {Landi}, R., {et~al.} 2010, \mnras, 408, 422

\bibitem[{{Stephen} {et~al.}(2006){Stephen}, {Bassani}, {Malizia}, {Bazzano},
  {Ubertini}, {Bird}, {Dean}, {Lebrun}, \& {Walter}}]{stephen06}
{Stephen}, J.~B., {Bassani}, L., {Malizia}, A., {et~al.} 2006, \aap, 445, 869

\bibitem[{{Stephen} {et~al.}(2005){Stephen}, {Bassani}, {Molina}, {Malizia},
  {Bazzano}, {Ubertini}, {Dean}, {Bird}, {Lebrun}, {Much}, \&
  {Walter}}]{stephen05}
{Stephen}, J.~B., {Bassani}, L., {Molina}, M., {et~al.} 2005, \aap, 432, L49

\bibitem[{{Ubertini} {et~al.}(2003){Ubertini}, {Lebrun}, {Di Cocco}, {Bazzano},
  {Bird}, {Broenstad}, {Goldwurm}, {La Rosa}, {Labanti}, {Laurent}, {Mirabel},
  {Quadrini}, {Ramsey}, {Reglero}, {Sabau}, {Sacco}, {Staubert}, {Vigroux},
  {Weisskopf}, \& {Zdziarski}}]{Ubertini:2003}
{Ubertini}, P., {Lebrun}, F., {Di Cocco}, G., {et~al.} 2003, \aap, 411, L131

\bibitem[{{Vollmer}(2009)}]{vollmer09}
{Vollmer}, B. 2009, VizieR Online Data Catalog, 8085, 0

\bibitem[{{White} {et~al.}(2000){White}, {Giommi}, \& {Angelini}}]{white00}
{White}, N.~E., {Giommi}, P., \& {Angelini}, L. 2000, VizieR Online Data
  Catalog, 9031, 0

\bibitem[{{Winkler} {et~al.}(2003){Winkler}, {Courvoisier}, {Di Cocco},
  {Gehrels}, {Gim{\'e}nez}, {Grebenev}, {Hermsen}, {Mas-Hesse}, {Lebrun},
  {Lund}, {Palumbo}, {Paul}, {Roques}, {Schnopper}, {Sch{\"o}nfelder},
  {Sunyaev}, {Teegarden}, {Ubertini}, {Vedrenne}, \& {Dean}}]{Winkler:2003}
{Winkler}, C., {Courvoisier}, T.~J.-L., {Di Cocco}, G., {et~al.} 2003, \aap,
  411, L1

\bibitem[{{Wright} {et~al.}(1994){Wright}, {Griffith}, {Burke}, \&
  {Ekers}}]{wright94}
{Wright}, A.~E., {Griffith}, M.~R., {Burke}, B.~F., \& {Ekers}, R.~D. 1994,
  \apjs, 91, 111

\bibitem[{{Wright} {et~al.}(2010){Wright}, {Eisenhardt}, {Mainzer}, {Ressler},
  {Cutri}, {Jarrett}, {Kirkpatrick}, {Padgett}, {McMillan}, {Skrutskie},
  {Stanford}, {Cohen}, {Walker}, {Mather}, {Leisawitz}, {Gautier}, {McLean},
  {Benford}, {Lonsdale}, {Blain}, {Mendez}, {Irace}, {Duval}, {Liu}, {Royer},
  {Heinrichsen}, {Howard}, {Shannon}, {Kendall}, {Walsh}, {Larsen}, {Cardon},
  {Schick}, {Schwalm}, {Abid}, {Fabinsky}, {Naes}, \& {Tsai}}]{wright10}
{Wright}, E.~L., {Eisenhardt}, P.~R.~M., {Mainzer}, A.~K., {et~al.} 2010, \aj,
  140, 1868

\bibitem[{{Zombeck}(1990)}]{zombeck90}
{Zombeck}, M.~V. 1990, {Handbook of space astronomy and astrophysics}, ed.
  {Zombeck, M.~V.}

\end{thebibliography}

\end{document}